\documentclass{aastex63}

\received{xxx}
\revised{xxx}
\accepted{xxx}
\submitjournal{ApJ}

\shorttitle{Radio flares of Cygnus X-3}
\shortauthors{Egron et al.}
\begin{document}

\title{Investigating the mini and giant radio flare episodes of Cygnus X-3
%The orbital modulation of Cygnus X-3 emission during the mini and giant radio flare episodes
%\footnote{Released on June, 10th, 2019}
}

\correspondingauthor{Elise Egron}
\email{elise.egron@inaf.it}

\author[0000-0002-1532-4142]{Elise Egron}
\affiliation{INAF-Osservatorio Astronomico di Cagliari, Via della Scienza 5, 09047 Selargius, Italy}

\author[0000-0002-4590-0040]{Alberto Pellizzoni}
\affiliation{INAF-Osservatorio Astronomico di Cagliari, Via della Scienza 5, 09047 Selargius, Italy}

\author[0000-0001-7332-5138]{Simona Righini}
\affiliation{INAF, Istituto di Radio Astronomia di Bologna, Via P. Gobetti 101, 40129 Bologna, Italy}

\author{Marcello Giroletti}
\affiliation{INAF, Istituto di Radio Astronomia di Bologna, Via P. Gobetti 101, 40129 Bologna, Italy}

\author{Karri Koljonen}
\affiliation{Finnish Centre for Astronomy with ESO (FINCA), University of Turku, V\"{a}is\"{a}l\"{a}ntie 20, 21500 Piikki\"{o}, Finland}
\affiliation{Aalto University Mets\"ahovi Radio Observatory, Mets\"ahovintie 114, 
02540 Kylm\"al\"a, Finland}
%{Aalto University Mets\"{a}hovi Radio Observatory, PO Box 13000, FI-00076 Aalto, Finland}

\author{Katja Pottschmidt}
\affiliation{CRESST and NASA Goddard Space Flight Center, Astrophysics Science Division Code 661, Greenbelt, MD 20771, USA}
\affiliation{Center for Space Science and Technology, University of Maryland Baltimore County, 1000 Hilltop Circle, Baltimore, MD 21250, USA}

\author{Sergei Trushkin}
\affiliation{Special Astrophysical Observatory of the Russian Academy of Sciences, Niznij Arkhyz 369167, Russia}
\affiliation{Kazan Federal University, Kazan, 420008, Russia}

\author[0000-0001-7738-2244]{Jessica Lobina}
\affiliation{Dipartimento di Fisica, Universit\`{a} degli Studi di Cagliari, SP Monserrato-Sestu, KM 0.7, I-09042 Monserrato, Italy}

\author[0000-0001-7397-8091]{Maura Pilia}
\affiliation{INAF-Osservatorio Astronomico di Cagliari, Via della Scienza 5, 09047 Selargius, Italy}

\author{Joern Wilms}
\affiliation{Dr. Karl-Remeis-Sternwarte and Erlangen Centre for Astroparticle Physics (ECAP), Friedrich Alexander Universit\"{a}t Erlangen-N\"{u}rnberg, Sternwartstr. 7, 96049 Bamberg, Germany}

\author{St\'{e}phane Corbel}
\affiliation{Lab AIM, CEA/CNRS/Universit\'e Paris-Saclay, Universit\'e de Paris, F-91191 Gif-sur-Yvette, France}
\affiliation{Station de Radioastronomie de Nan\c{c}ay, Observatoire de Paris, PSL Research University, CNRS, Univ. Orl\'{e}ans, 18330 Nan\c{c}ay, France}

\author[0000-0003-2538-0188]{Victoria Grinberg}
\affiliation{Institut f\"{u}r Astronomie und Astrophysik, Universit\"{a}t T\"{u}bingen, Sand 1, 72076 T\"{u}bingen, Germany}

\author[0000-0001-5126-1719]{Sara Loru}
\affiliation{INAF-Osservatorio Astrofisico di Catania, Via S.Sofia 78, 95123 Catania, Italy}

\author{Alessio Trois}
\affiliation{INAF-Osservatorio Astronomico di Cagliari, Via della Scienza 5, 09047 Selargius, Italy}

\author[0000-0002-4151-4468]{J\'{e}rome Rodriguez}
\affiliation{Lab AIM, CEA/CNRS/Universit\'e Paris-Saclay, Universit\'e de Paris, F-91191 Gif-sur-Yvette, France}
         
\author{A. L\"ahteenm\"aki}
\affiliation{Aalto University Mets\"ahovi Radio Observatory, Mets\"ahovintie 114, 
02540 Kylm\"al\"a, Finland}
\affiliation{Aalto University Department of Electronics and Nanoengineering,  
P.O. BOX 15500, FI-00076 AALTO, Finland}

\author{M. Tornikoski}
\affiliation{Aalto University Mets\"ahovi Radio Observatory, Mets\"ahovintie 114, 
02540 Kylm\"al\"a, Finland}

\author{S. Enestam} 
\affiliation{Aalto University Mets\"ahovi Radio Observatory, Mets\"ahovintie 114, 
02540 Kylm\"al\"a, Finland}

\author{E. J\"arvel\"a}
\affiliation{Aalto University Mets\"ahovi Radio Observatory, Mets\"ahovintie 114, 
02540 Kylm\"al\"a, Finland}

%\collaboration{1}{(AAS Journals Data Scientists collaboration)}

%% Note that the \and command from previous versions of AASTeX is now
%% depreciated in this version as it is no longer necessary. AASTeX 
%% automatically takes care of all commas and "and"s between authors names.

%% AASTeX 6.3 has the new \collaboration and \nocollaboration commands to
%% provide the collaboration status of a group of authors. These commands 
%% can be used either before or after the list of corresponding authors. The
%% argument for \collaboration is the collaboration identifier. Authors are
%% encouraged to surround collaboration identifiers with ()s. The 
%% \nocollaboration command takes no argument and exists to indicate that
%% the nearby authors are not part of surrounding collaborations.

%% Mark off the abstract in the ``abstract'' environment. 
\begin{abstract}

The microquasar Cygnus X-3 underwent a giant radio flare in April 2017, reaching a maximum flux of $\sim 16.5$ Jy at 8.5 GHz. We present results from a long monitoring campaign carried out with Medicina at 8.5, 18.6 and 24.1 GHz, in parallel to the Mets\"ahovi radio telescope at 37 GHz, from 4 to 11 April 2017.
%Alternating the three frequencies every $\sim30$\,min during the observations that lasted for $3-11$\,h per day allowed us to investigate the evolution of the spectral index on very short timescales. 
We observe a spectral steepening from $\alpha = 0.2$ to 0.5 (with $S_{\nu} \propto \nu^{-\alpha}$) within 6\,h around the epoch of the peak maximum of the flare, and rapid changes in the spectral slope in the following days during brief enhanced emission episodes while the general trend of the radio flux density indicated the decay of the giant flare.
We further study the radio orbital modulation of Cyg X-3 emission associated with the 2017 giant flare and with six mini-flares observed in 1983, 1985, 1994, 1995, 2002 and 2016. 
The enhanced emission episodes observed during the decline of the giant flare at 8.5 GHz coincide with the orbital phase $\phi \sim 0.5$ (orbital inferior conjunction). 
%Instead, the light curves of the mini-flares observed at $15-22$ GHz are aligned in phase, peaking at $\phi \sim 0$, except for the 2016 light curve which is shifted of $\sim 0.5$ w.r.t. the other ones. 
On the other hand the light curves of the mini-flares observed at $15-22$ GHz peak at $\phi \sim 0$, except for the 2016 light curve which is shifted of 0.5 w.r.t. the other ones.
%We attribute the apparent phase shift to the different accretion states (the 2016 mini-flare occurred in the hypersoft X-ray state) and/or different jet geometries, whose precise knowledge is needed to implement barycentric timing corrections.
We attribute the apparent phase shift to the variable location of the emitting region along the bent jet. This might be explained by the different accretion states of the flaring episodes (the 2016 mini-flare occurred in the hypersoft X-ray state).
%timing correction for the motion and the variable location of the emitting region in different flaring episodes at different epochs

%%%\textbf{In the case of the giant flare, the enhanced emission episodes that were observed during the decline of the flare coincide with the orbital phase $\phi \sim 0.5$ (orbital inferior conjunction), whereas the 2016 mini-flare appear to be shifted in phase w.r.t. the other mini-flares. We attribute the apparent phase shift to the different accretion states (the 2016 mini-flare occurred in the hypersoft X-ray state) and/or different jet geometries, whose precise knowledge is needed to implement barycentric timing corrections.} 
%(barycentric corrections are needed to properly correct time delays arising from jet emission at different altitudes.}
%This can be explained by a different location of the emitting region.
%Our  timing  results  confirm  that  the  studied  mini-flares  occurred  during  different  accretionstates and/or different emission geometries.
%, a different value of the period derivative with respect to X-rays, and/or because the system’s radio light curves experienced phase drifts as recently seen in X-rays.
%This can be explained by variations in the orbital phase and location of the emitting region and/or a different value of the period derivative with respect to X-rays.

\end{abstract}

%% Keywords should appear after the \end{abstract} command. 
%% See the online documentation for the full list of available subject
%% keywords and the rules for their use.

%% From the front matter, we move on to the body of the paper.
%% Sections are demarcated by \section and \subsection, respectively.
%% Observe the use of the LaTeX \label
%% command after the \subsection to give a symbolic KEY to the
%% subsection for cross-referencing in a \ref command.
%% You can use LaTeX's \ref and \label commands to keep track of
%% cross-references to sections, equations, tables, and figures.
%% That way, if you change the order of any elements, LaTeX will
%% automatically renumber them.
%%
%% We recommend that authors also use the natbib \citep
%% and \citet commands to identify citations.  The citations are
%% tied to the reference list via symbolic KEYs. The KEY corresponds
%% to the KEY in the \bibitem in the reference list below. 

%%\label{sec:intro}

\section{Introduction}

Discovered in 1966 \citep{Giacconi_1967}, Cygnus X-3 is a unique and enigmatic X-ray binary system that consists in a low-mass compact object \citep[$M \sim 2.4^{+2.1}_{-1.1}\,\,\mathrm{M_{\sun}}$;][]{Zdziarski_2013} and a Wolf-Rayet star \citep{vanKerkwijk_1992, vanKerkwijk_1996, koljonen_2018b}. The exact nature of the compact object is still unknown, but a black hole is favoured according to its X-ray and radio properties \citep{Hjalmarsdotter_2008, Hjalmarsdotter_2009, Szostek_2008, Koljonen_2010}. Contrary to other high-mass X-ray binaries (HMXBs), Cyg X-3 has a very short binary period $P_\mathrm{orb} \sim 4.8$\,h,
%(Parsignault et al. 1972), 
which implies a very tight orbit ($\sim 3 \times 10^{11}$\,cm) and strong interactions between the compact object and the Wolf-Rayet wind \citep[$\dot{M} \sim 10^{-5}$ M$_{\sun}$\,yr$^{-1}$, $v_\mathrm{wind} \sim 1000$ km\,s$^{-1}$;][]{vanKerkwijk_1996,koljonen_2018b}.
The compact object appears to be totally enshrouded in the wind of the donor star.

Located in the Galactic plane at a distance of $\sim 7.4 \pm 1.1$ kpc \citep{McCollough_2016}, Cyg X-3 is the brightest microquasar at radio wavelengths. Although being most of the time in the quiescent state ($\sim 100$ mJy), Cyg X-3 is characterized by a highly variable radio emission spanning from $1-30$ mJy in the quenched radio state \citep{Waltman_1996} up to 20 Jy during the flaring states \citep{Waltman_1995, Mioduszewski_2001, Miller-Jones_2004, Corbel_2012}. 
Differently from what is observed in other microquasars, singular giant radio flares correspond to the transition from the ultra-soft  X-ray state \citep[so-called hypersoft state;][]{Szostek_2008, Koljonen_2010, Koljonen_2018} to a harder state.
%hypersoft X-ray state to a harder state \citep{Koljonen_2010, Koljonen_2018}. 
The hypersoft state corresponds to unusual very soft X-ray spectra, associated with a quenched radio emission during which there is no or very faint radio emission that can last several weeks preceding the giant radio flare.
%for several weeks. 
The rise of the giant flares usually lasts for about 3 days while the decay is visible over $\sim 10-30$ days \citep{Trushkin_2017}. Relativistic jets are clearly resolved during these episodes using the Very Large Array (VLA), 
the Very Long Baseline Array (VLBA) and the European Very Long Baseline Interferometry Network (e-EVN) \citep{Marti_2001, Mioduszewski_2001, Miller-Jones_2004, Tudose_2007}. 
Mini-flares were detected during the quiescent and quenched radio states. They are characterized by low-flux amplitudes (up to 1 Jy) and duration less than a day
% a very short flare duration of the order of a few hours 
\citep{Waltman_1994}. They are most likely associated with the compact jet emission \citep{Newell_1998, Egron_2017}. 
%(MORE INFO ABOUT EXTENDED AND COMPACT JETS WILL BE ADDED...)

%Gamma, X-ray, IR obserations, orbital modulations.
A flux modulation, taken to represent the orbital period of the binary system, was first observed in the X-rays, \citep{Parsignault_1972, Canizares_1973}
%Sanford \& Hawkins 1972, 
and subsequently in the infrared \citep{Becklin_1973}. 
The X-ray minima occur at the superior conjunction, when the compact object lies behind the Wolf Rayet star. 
%\citet{Zdziarski_2012} studied the dependence of the X-ray orbital modulation with energy and X-ray spectral states. 
The X-ray modulation depth decreases with increasing energy above 5 keV, which is interpreted  as bound-free absorption and Compton scattering in the stellar wind of the donor \citep{Zdziarski_2012}. In this scenario, the X-ray minima correspond to the highest optical depth at the superior conjunction. A decrease of the optical depth below 3 keV is instead attributed to the re-emission of the absorbed continuum by the wind in soft X-ray lines \citep{Zdziarski_2012}.
The X-ray orbital period increases with time \citep[$\dot{P}/P \sim 1 \times 10^{-6}$ yr$^{-1}$;][]{vanderklis_1981, Kitamoto_1987, Kitamoto_1995, Singh_2002}. This slow-down is ascribed to the loss of angular momentum through the mass-loss of the stellar wind \citep{Davidsen_1974}. 

Cyg X-3 was the first microquasar detected in the gamma-rays above 100 MeV with Fermi/LAT and AGILE \citep{Fermi_2009, Tavani_2009}.
%, opening new areas to the jet formation and the connection with the radio emission \citep{Dubus_2010}. 
Inverse Compton scattering of soft photons from the Wolf-Rayet companion on relativistic electrons in the jet naturally explains the gamma-ray emission \citep{Dubus_2010}. This high-energy emission is associated with transitions into and out of the quenched radio state, probably due to shocks forming at various distances along the jet \citep{Corbel_2012, Koljonen_2018}. Bright gamma-ray emission is also detected during the mini-flares, when shocks occur closer to the core of the jet, hence deeper in the Wolf-Rayet wind.
%close to the Wolf-Rayet star. 
%The energy density in seed photons instead decreases when shocks occur  far downstream, reducing the inverse-Compton scattering and so the gamma-ray emission during brighter radio flares.
%Particle acceleration to very high energies
% and therefore to changes in jet efficiency.
% implying a connection to the accretion process, and also that the γ-ray activity is related to the level of radio flux (and possibly shock formation), strengthening the connection to the relativistic jets. 
The gamma-ray emission was found to be strongly modulated at the orbital period of the system \citep{Fermi_2009, Zdziarski_2018}, but shifted of a half orbital period with respect to the X-ray modulation.
%However, the phases of gamma-ray flux maxima and minima are close to the phases of  X-ray flux minima and maxima, respectively. The gamma-ray modulation.
%The gamma-ray modulation is explained by the anisotropic Inverse Compton scattering of soft photons from the Wolf-Rayet star by relativistic electrons in the jet along the orbit \citep{Dubus_2010}.
A jet inclination of $\sim 30^\circ$ with respect to the orbital plane is required in order to obtain good fits to the gamma-ray modulation \citep{Dubus_2010,Zdziarski_2018}.
%For a perpendicular jet, the maximum of that emission would be at $\phi = 0$. Since it is not the case, a jet inclination of $\sim 30^\circ$ is required. 
%The calculations were updated using much longer duration data in \citet{Zdziarski_2018}.
An inclined jet is likely to undergo precession and strongly affect the gamma-ray modulation shape at different epochs. However, up to date there is no evidence for such a jet precession in gamma-rays. Instead, a jet precession was clearly detected during giant radio flares at much larger distances from the compact object \citep{Mioduszewski_2001,Miller-Jones_2004}.
%The jet precession is expected to strongly affect the gamma-ray modulation shape at different epochs.}
%The maximum occur when stellar photons are backscattered towards the observer. 
%upscattering of soft photons from the Wolf-Rayet star by relativistic electrons in the jet 

A modulation of the radio flux at the orbital period has been claimed in the HMXB Cyg X-1 \citep{Fender_1997IAUC, Pooley_1999}. This variable absorption of the flux density has been attributed to the stellar wind, as the black-hole jet orbits around the OB companion star \citep{Brocksopp_2002, Szostek_2007}. A similar orbital modulation was also expected in Cyg X-3. 
%In fact, a
%
A 4.8-h orbital modulation has been highlighted at radio wavelengths during low-level radio flares from Cyg X-3 \citep{Molnar_1984}, but no orbital modulation was later confirmed.
% mostly because of the orbital period being exactly 1/5 of a day. 
%However, a careful study of the archival radio emission revealed the orbital modulation also in the radio (Zdziarski et al., subm.). 
% (CITE ZDZIARSKI ET AL. SUBMITTED HERE ?).
%Gamma-ray (Atel 10252 Loh, Fermi/LAT) ; (ATels 10138/10179 Piano, AGILE) and radio observations (Atel 10126 ; Trushkin) on Cyg X-3 predicted a forthcoming giant flare only a few months after the last spectacular event in September 2016 (Egron et al. 2017, etc....). 
A pronounced modulation of the radio emission has been recently detected by \citet{Zdziarski_2018}, who performed a careful study of the archival radio emission at 15 GHz.
%, based on the radio flux level and the X-ray spectral state.
The amplitude of the modulation is found to depend on both the X-ray spectral state and the radio flux, changing from $2.5\%$ to $10\%$.
The radio orbital modulation is well modelled by free–free absorption by the stellar wind.

VLBI observations performed at 22 GHz during the 2016 mini-flare showed a short time-scale variability, with the evidence of two peaks at different flux densities \citep{Egron_2017}. At a first glance, the difference between the two peaks corresponds to the orbital period of Cyg X-3.
In order to investigate possible orbital modulation effects during mini-flare events, we collected previous mini-flare observations \citep[above 100\,mJy and below 1\,Jy;][]{Waltman_1996} carried out at 15 and 22 GHz with interferometric antennas.
The motivation for
%
%In order to investigate a possible orbital modulation effects during the mini-flare events, we collected previous mini-flare observations carried out at high radio frequencies with interferometric antennas. The advantage of
% 
using such observations is their duration (usually several continuous hours), which is of utmost importance to provide a dense flux coverage of at least a whole orbital period of the binary system. 
%
%Furthermore, the modulation amplitude is most likely larger at higher frequencies, as observed in Cyg X-1 \citep{Brocksopp_2002}. 
We excluded data at other frequencies since they are not enough sampled and therefore not suitable for a statistic analysis of light-curves when considering them alone.

In this paper, we present the monitoring of the 2017 April giant flare carried out with the Medicina 32-m and the Mets\"ahovi 13.7-m radio telescopes. We study the evolution of the radio flux and spectral index during this event. In the frame of the presently limited and sparse samples preventing an extensive orbital timing analysis, we provide evidence of 
%we investigate the
radio orbital modulation during specific episodes of Cyg X-3, the giant flare and mini flares, where the jet characteristics and morphologies are very different.
%We investigated the radio orbital modulation during specific episodes of Cyg X-3  the giant and mini radio flares
%%%We then investigate the orbital modulation during the giant flare and during the mini flares of Cyg X-3, where the jet characteristics and morphologies are very different. 

%Note that we read the paper from \citet{Zdziarski_2018} who performed a careful study of the archival radio emission and revealed the orbital modulation in the radio while our paper was in preparation. Our study confirms the orbital modulation in radio using another method, but we propose alternative interpretations of the results. 

\section{Radio data}

\subsection{Medicina multi-frequency observations during the 2017 giant flare}

\begin{figure*}
   \centering
   \includegraphics[width=0.80\textwidth]{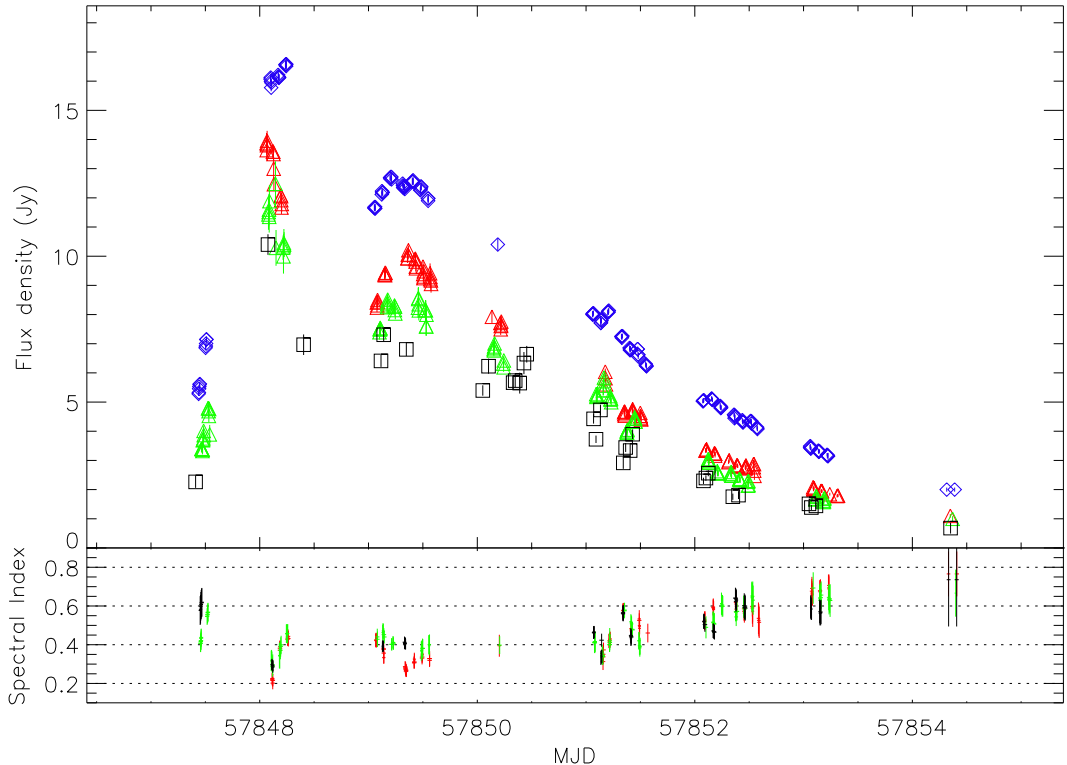}
%    \vspace{0.8cm}
    \caption{Top panel: Medicina observations performed at 8.5, 18.6 and 24.1 GHz (blue diamonds, red and green triangles, respectively) and MRO observations carried out at 37 GHz (black squares) during the giant flare from 2017 April 4 to April 11. Bottom panel: Evolution of the spectral index $\alpha$ between 8.5 and 18.6 GHz (red), between 8.5 and 24.1 GHz (green), and between 8.5 and 37 GHz (black).}
    \label{fig:flux-alpha-2017}
\end{figure*}

%Thanks to the RATAN daily monitoring of Cyg X-3 \citep{Trushkin_2017}, 
Following the detection of the start of the flare in Cyg X-3 by the RATAN \citep{Trushkin_2017}, we triggered a Target-of-Opportunity program with the Medicina radio telescope (32-m; \textit{www.med.ira.inaf.it}) in order to follow the evolution of the radio emission during the whole duration of the giant flare. The frequency agility offered by Medicina (i.e. the change of the observing receiver/frequency requiring at most a few minutes), allowed us to carry out observations at 8.5, 18.6 and 24.1 GHz (typically 30 min each) from the $4^{th}$ to the $11^{th}$ April 2017. These long sessions lasting from 3 to 11 h per day were aimed at tracking possible changes in the radio spectral index.
We performed On-The-Fly cross-scans in RA and DEC directions, setting a bandwidth of 680 MHz. Scans were performed along a length of $0.6^{\circ}$, at a velocity of 2.4'/sec at 8.5\,GHz. We selected a bandwidth of 1200 MHz, scans of $0.2^{\circ}$ in length and a scan velocity of 0.8'/sec in \textit{K}-band (18.6 and 24.1 GHz). 
Data calibration was carried out through the observation of NGC\,7027 at each frequency, within 1\,h from Cyg X-3 acquisitions. 
We extrapolated the calibrator flux density according to \citet{Ott_1994}:
%\citet{Perley_2013}:
$5.764\pm 0.005$ Jy at 8.5 GHz, $5.508\pm 0.009$ Jy at 18.6 GHz, and $5.367\pm 0.015$ Jy at 24.1 GHz.

The calibration procedure included the corrections for the frequency-dependent gain curves, plus the compensations for the pointing offset measured on each scan. Corrections for  atmospheric opacity were applied to the \textit{K}-band measurements only.
%We applied gain curve, pointing offset optimization, opacity and atmospheric corrections to the \textit{K}-band measurements. 
In order to improve the accuracy on the flux density measurements, we carried out additional observations on the bright HII region DR21 ($F \sim 20$ Jy at 18.6 GHz) because of its location very close to Cyg X-3 (angular separation $\sim 1.8^\circ$). It allowed us to apply additional gain curve corrections, in particular at low elevations, where the standard curve is less effectual and is characterised by larger residuals. We rejected the \textit{K}-band data acquired under unsuitable weather conditions 
(K-band requires a uniform sky opacity as far as possible and no rain).
We estimate the final accuracy of our measurements to be $\sim 5\%$ in \textit{K}-band, while we obtain $\sim 3\%$ errors at 8.5 GHz. The resulting light curve is presented in Figure~\ref{fig:flux-alpha-2017} and the associated data are shown in Appendix in  Tables~\ref{tab:A1}, \ref{tab:A2} and \ref{tab:A3}.

\subsection{Mets\"ahovi radio telescope monitoring during the 2017 giant flare}

With the onset of the Cyg X-3 outburst in April 2017, we started monitoring observations with the 13.7\,m diameter Mets\"ahovi radio observatory (MRO) at 37 GHz. The observations were made in dual beam switching mode, alternating the source and the sky in each feed horn. A typical integration time to obtain one flux density data point was between 1200 and 1400\,s. The detection limit of the telescope at 37 GHz is on the order of 0.2\,Jy under optimal conditions. Data points with a signal-to-noise ratio $<$\,4 are handled as non-detections.

The flux density scale is set by observations of DR\,21. Sources NGC\,7027, 3C\,274, and 3C\,84 are used as secondary calibrators. A detailed description of the data reduction and analysis is given in \citet{Teraesranta_1998}. The error estimate in the flux density includes the contribution from the measurement rms and the uncertainty of the absolute calibration. The data are shown in Figure~\ref{fig:flux-alpha-2017} together with the Medicina multi-frequency observations, and repored in Annex in Table~\ref{tab:A4}.

\subsection{Interferometric data during the mini flares}

We used interferometric observations from the literature that contained mini-flare observations of Cyg X-3: 
Very Large Array (VLA)  data from 1983 September 17-18 \citep{Molnar_1984}, 
VLBI data from 1985 February 5 \citep{Molnar_1988}, 
Ryle telescope data obtained on 1994 June 3 \citep{Waltman_1996},
Very Long Baseline Array (VLBA) data from 1995 May 7 \citep{Newell_1998},
Ryle telescope and VLA observations from 2002 January 25 \citep{Miller-Jones_2009},
and more recently VLBI observations corresponding to the mini-flare observed a few days before the onset of the giant flare on 2016 September 1 \citep{Egron_2017}.
%data set obtained in 1983, 1985, 1995 and more recently on 2016.

VLA obervations were carried out in A configuration on 1983 September 17, 18 from 1:30 to 7:30 UT (both days) at 1.3, 2, 6 and 20\,cm. For our study, we restricted our analysis to the 2\,cm data (15 GHz).
A composite array of 5 VLBI antennas was used to perform observations of Cyg X-3 at 22 GHz on 1985 February 5. Due to some telescope problems, only 4 antennas were employed during both sets of observations, which lasted for $\sim 7$ h.
Observations of Cyg X-3 were carried out at 15 GHz with the Ryle telescope in 1994 and 1995, during gaps between the prime observing targets.
The VLBA observations performed on 1995 May 7 from 08:40 to 17:45 UT were conducted at 15.3 GHz using all 10 antennas. The data were correlated at the VLBA correlator in Socorro, New Mexico. 
Cygnus X-3 was observed on 2002 January 25 for 8\,h with the VLA in A-configuration.  The VLA was
split into two subarrays in order perform simultaneous observations at 14.94 GHz and at 43.34 GHz.
The source was also observed at 15 GHz by the Ryle telescope, overlapping the VLA observations for 1.6\,h at the beginning of the VLA run.
Recent VLBI observations were conducted on 2016 September 1 at 22 GHz with the SRT (64\,m, Italy), Medicina (32\,m, Italy), Noto (32\,m, Italy), Torun (32\,m, Poland), Yebes (40\,m, Spain) and Onsala (20\,m, Sweden) for approximately 12 h. The data were processed with the DiFX correlator \citep{Deller_2011} installed and operated in Bologna, Italy. 

\begin{figure*}
\gridline{\fig{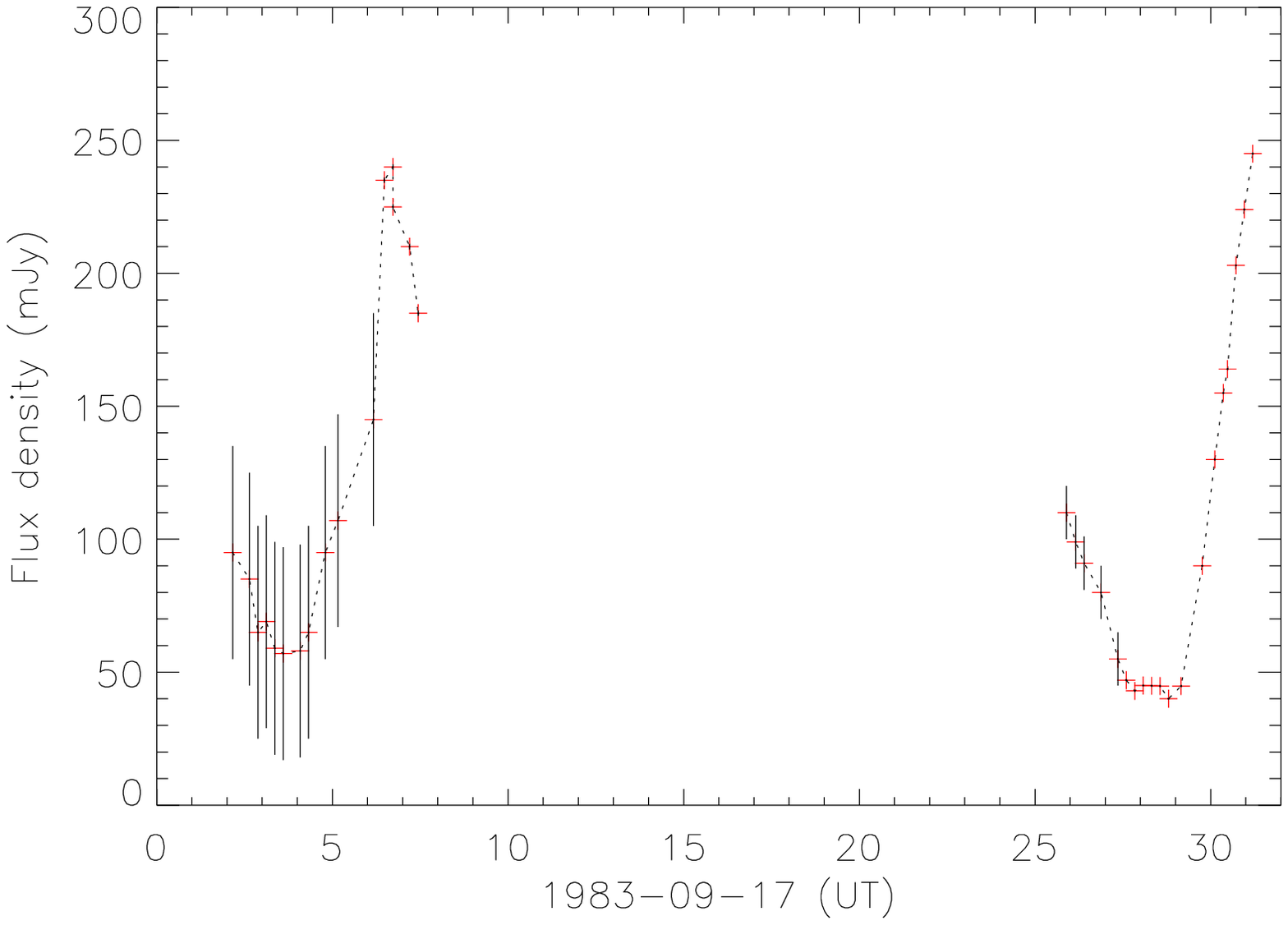}{0.3\textwidth}{(a)}
          \fig{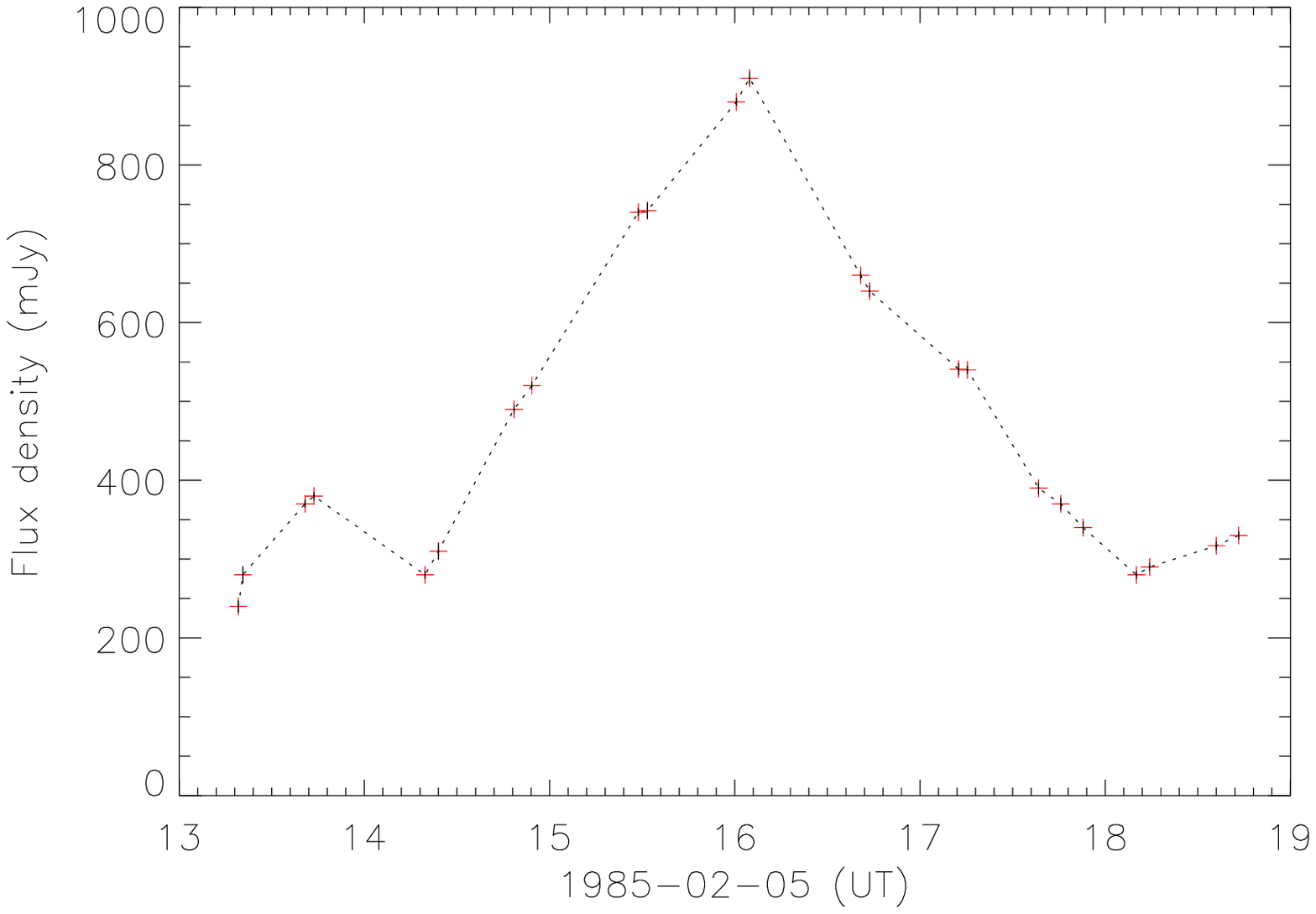}{0.3\textwidth}{(b)}
          \fig{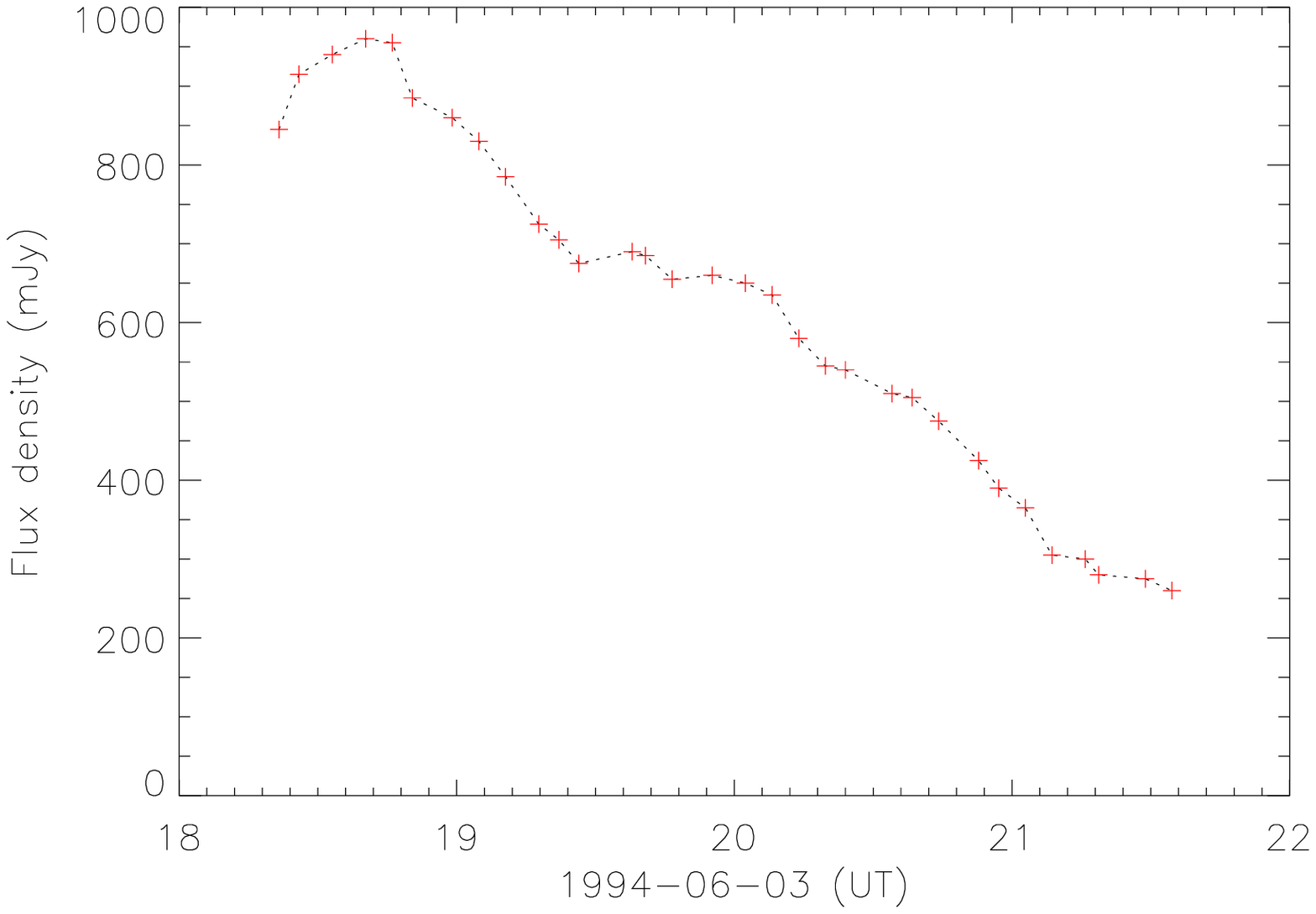}{0.3\textwidth}{(c)}
          }
\gridline{\fig{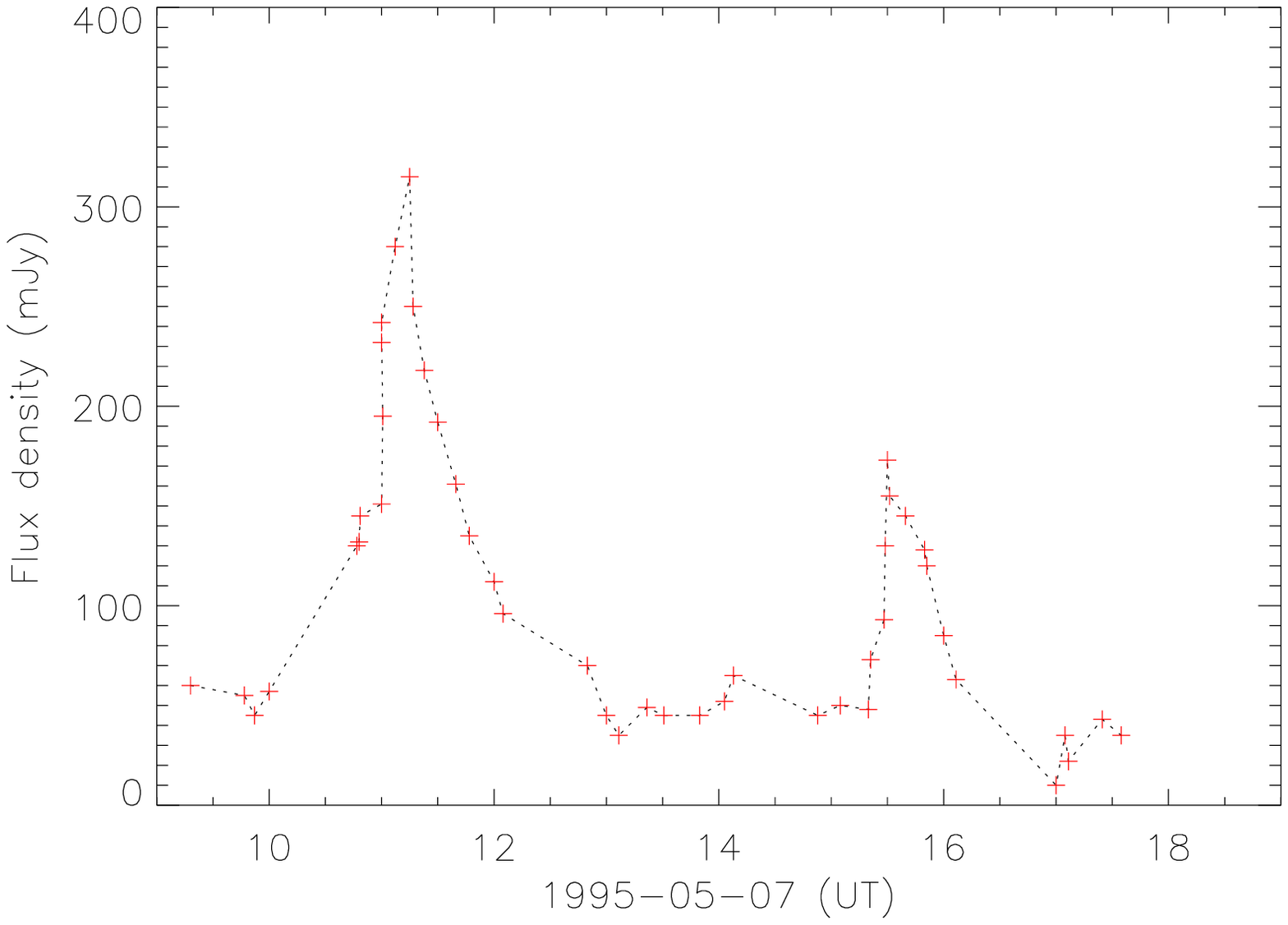}{0.3\textwidth}{(d)}
          \fig{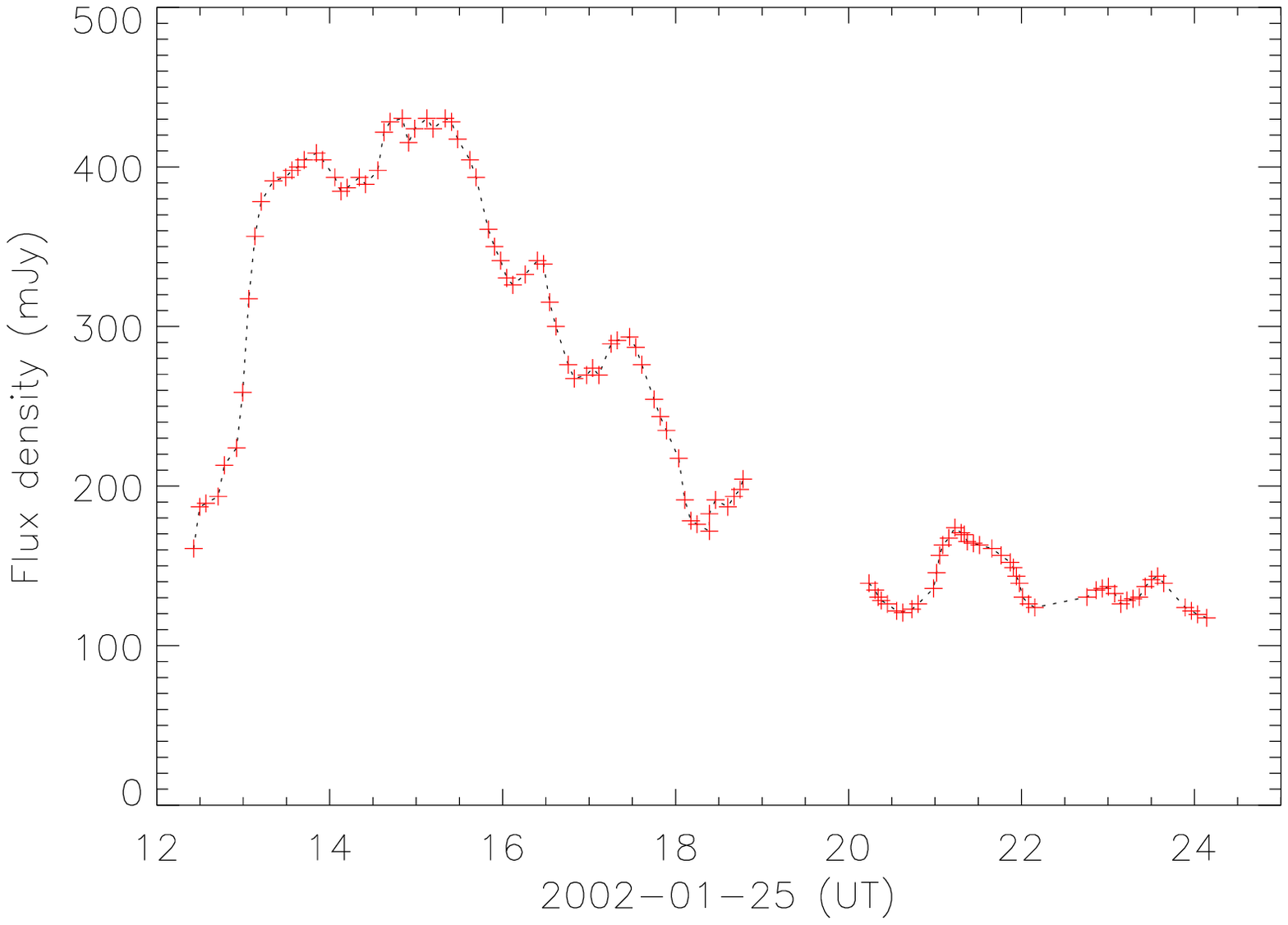}{0.3\textwidth}{(e)}
	  \fig{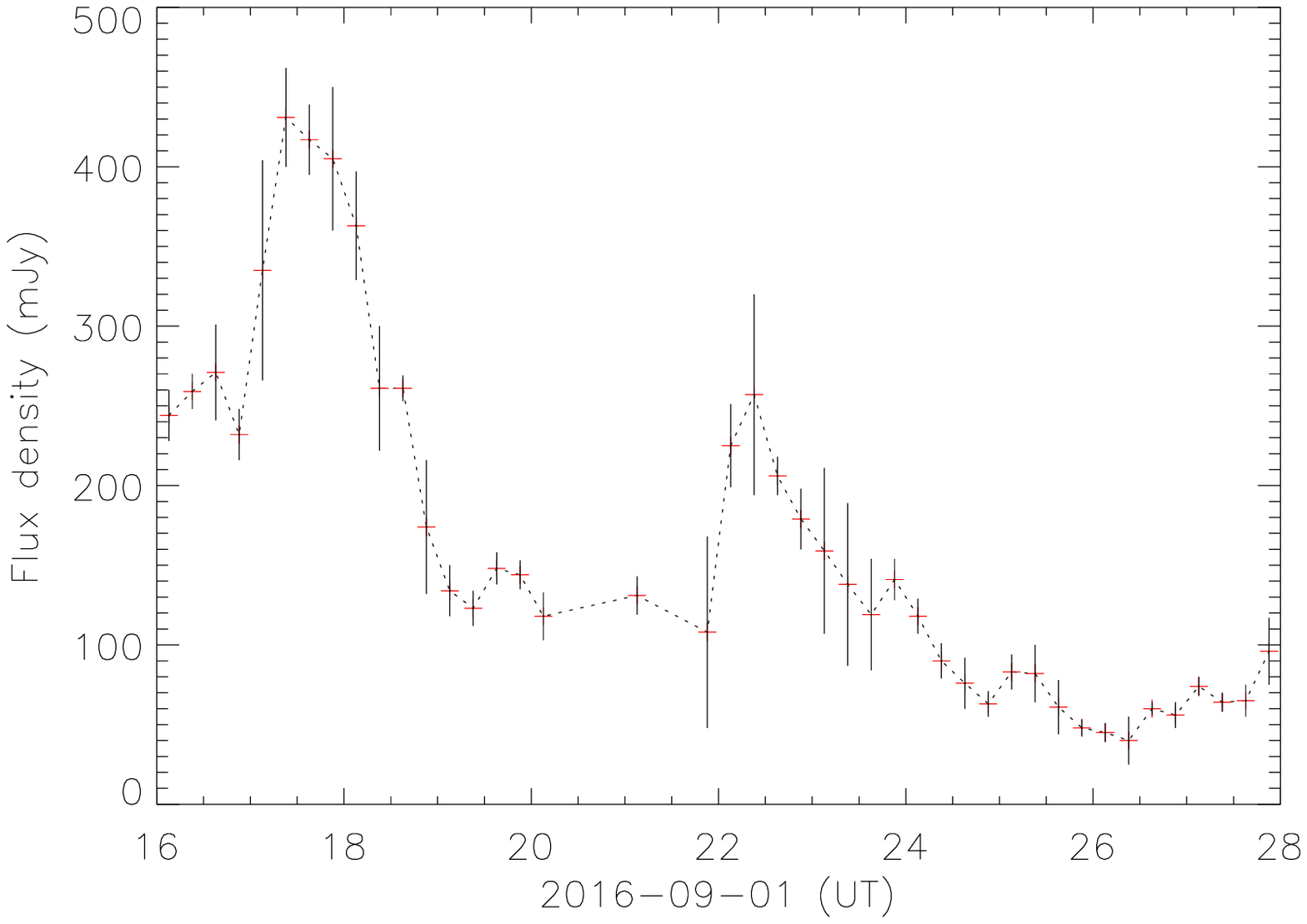}{0.3\textwidth}{(f)}
          }
\caption{Light curves of Cyg X-3 obtained during mini-flares on
a) 1983 September 17-18 (MJD 45594, 45595) at 15 GHz \citep{Molnar_1984} with the VLA,
b) 1985 February 5 (MJD 46101) at 22 GHz \citep{Molnar_1988} with the VLBI,
c) 1994 June 3 (MJD 49506) at 15 GHz \citep{Waltman_1996} with the Ryle telescope,
d) 1995 May 7 (MJD 49844) at 15 GHz \citep{Newell_1998} with the VLBA,
e) 2002 January 25 (MJD 52299) at 15 GHz with the Ryle telescope and the VLA \citep{Miller-Jones_2009},
f) 2016 September 1 (MJD 57632) at 22 GHz \citep{Egron_2017} with the VLBI.
\label{fig:6miniflares}}
\end{figure*}

We did not re-analyse the data, but used the resulting light curves presented in the different papers to perform our study. 
The associated light-curves are shown in Figure~\ref{fig:6miniflares}.
%{fig:lc-vla-17sept1983}$-$\ref{fig:lc-vlbi-1sept2016}.
%~\ref{fig:lc-vlbi-5feb1985},~\ref{fig:lc-ryle-3june1994},~\ref{fig:lc-vlba-7may1995},~\ref{fig:lc-vla-ryle-25 jan2002}, %and ~\ref{fig:lc-vlbi-1sept2016}.

%observed a flare of peak flux #0.95 Jy (at 22 GHz) in 1985 February using a composite array comprising five VLBI antennas (although telescope problems reduced this to four during both sets of observations). The observations covered the complete evolution of a flare which lasted for approximately 7 h

\section{The 2017 giant flare}

%\subsection{Evolution of the spectral index}

Radio flares are caused by synchrotron radiation from accelerated relativistic electrons suffering radiative, adiabatic and energy-dependent loss mechanisms. Particles can be accelerated in 1)  discrete plasmoids evolving from optically thick to optically thin as they move outward from the core and fastly expand \citep{van-der-Laan_1966, Hjellming_1988, Atoyan_1999}, 2) internal shocks in the jet \citep{Fender_2004, Lindfors_2007, Miller-Jones_2009}, or 3) magnetic reconnection in the relativistic plasma with a relatively high magnetization \citep{Guo_2014, Sironi_2014, Sironi_2016}.

During the April 2017 giant flare, the flux density of Cyg X-3 increased expeditiously to reach a peak maximum of 16.6 $\pm 0.3$ Jy at 8.5 GHz on April 5 (MJD 57848), as shown in Figure~\ref{fig:flux-alpha-2017}.
The radio emission observed at 18.6, 24.1 and 37 GHz followed a similar trend apart from the
8.5 GHz flux density, which was still rising on April 5 while the flux were already weakening at higher frequencies.
Figure~\ref{fig:flux-alpha-2017} provides evidence of small and short-duration flux fluctuation at different frequencies during the general declining trend of the giant flare on April 6 and April 8. 
A frequency-dependent slight delay between the different light curves has been observed for instance in \citet{Molnar_1984,Molnar_1988}. In our data, the above increases of flux density seem to appear firstly at higher frequency (37 GHz, then 24.1, 18.6 and 8.5 GHz), but the presence of a significant delay cannot be firmly confirmed due to the lack of perfectly simultaneous and continuous observations.
%The resulting light curve is presented in Figure~\ref{fig:flux-alpha-2017}.

We studied the evolution of the spectral index $\alpha$ (with $S_{\nu} \propto \nu^{-\alpha}$) by considering all pairs of flux density measurements available at low (8.5 GHz) and high frequencies (18.6, 24.1 and 37 GHz) within 1\,h from each other. Spectral index errors were derived from error propagation of flux density errors for each pair. The time tag of the reported spectral index values corresponds to the mid time between the epochs of each flux density measurement pair. The error bars on the x-axis (time) reflects the epoch separation for each pair of flux density measurements. 
As shown in Figure~\ref{fig:flux-alpha-2017}, we observed a spectral steepening from $\alpha = 0.22 \pm 0.05$ to $0.45 \pm 0.05$ between 8.5 GHz and 18.6 GHz, and from $\alpha = 0.3 \pm 0.1$ to $0.46 \pm 0.05$ between 8.5 GHz and 24.1 GHz within $\sim 6$\,h at the time of the peak maximum of the flare (April 5). The spectral index then continuously varied in the following days. In particular it flattened then steepened from $\alpha \sim 0.3$ to $\sim 0.4$ on MJD 57849 and from $\alpha \sim 0.3$ to $\sim 0.6$ on MJD 57851 when the radio flux density suddenly increased while the general trend indicated a flare decay. At the end of the giant flare ($F_{8.5 \mathrm{GHz}} < 5$ Jy), the spectral index reached a value of $\sim 0.7$.

Here, as well as in \citet{Egron_2017}, it has been shown that the radio spectral index varies in less than 6\,h at the time of the flare peak.
% (Figure~\ref{fig:flux-alpha-2017}), 
The values of the spectral index are very similar to the ones associated with previous giant flares
\citep[][and references therein]{Miller-Jones_2004}. However, the long monitoring performed with Medicina during the 2017 giant flare shows for the first time additional spectral variations (flattening then steepening) at the time of fast and short-duration increases of flux density during the decaying flare. 

\section{Orbital modulation}

%The orbital period of Cyg X-3 inferred from X-ray data slightly increases with time, likely due to the loss of the angular momentum through wind mass-loss (Davidsen \& Ostriker 1974).
We searched for orbital modulation in the radio emission during the giant and mini-flare episodes, which are associated with extended and compact jets, respectively. 
A modulation of the radio emission from the jet is expected by variable free-free absorption in the wind from the companion star along the orbit \citep{Molina_2019}.
% varying with the orbital phase.
We considered the recent quadratic ephemeris from \citet{Bhargava_2017}, based on 45 years of X-ray data obtained from the archives of the \textit{Einstein Observatory, EXOSAT, Ginga, RXTE} and the recent \textit{AstroSAT} data:
%, \textbf{all corrected to the solar system barycenter}:
% 
\begin{equation}\label{}
T_{n} = T_{0} + P_{0}n + \frac{1}{2} \dot{P}P_{0}n^{2},
\end{equation}
where $T_{0} = 40949.384$ (MJD) is the epoch at the superior conjunction ($\phi = 0$), $P_{0} = 0.19968476(3)$\,d the period at $T_{0}$, $\dot{P} = (5.42 \pm 0.02) \times 10^{-10}$ the period derivative, and $n$ the orbit number.
This recent ephemeris is based on barycentre-corrected X-ray light curves and given in Terrestrial Time MJD, 
%based on X-ray data with barycentric correction 
as mentioned in \citet{Zdziarski_2018}.
% (presumably related to the superior conjunction). 
%The period and period derivative were determined with a much better accuracy than previously \citep{Singh_2002}.

\begin{figure*}
   \centering
   \includegraphics[width=0.62\textwidth]{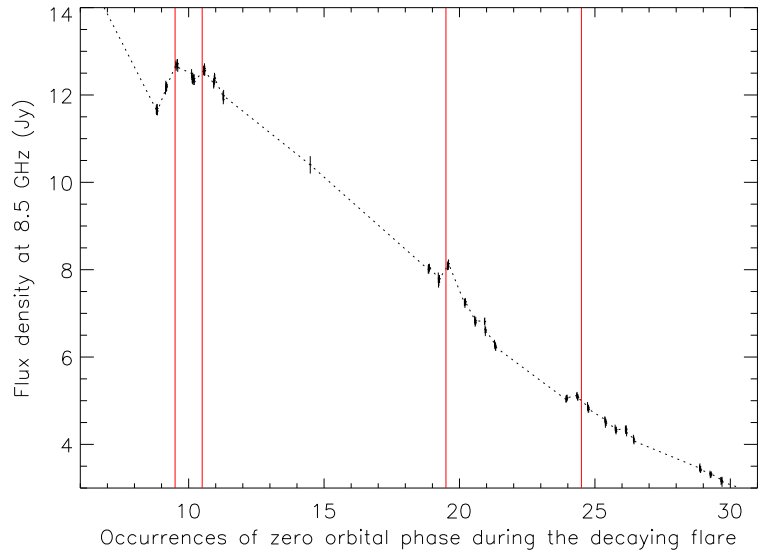}
    \caption{Medicina flux density at 8.5 GHz during the decaying giant flare in April 2017 as a function of the occurrences of zero orbital phase since the $84621.0^{th}$ orbital period of Cyg X-3. 
% during the decaying of the flare.
% at the moment of quick variations of the flux density. 
%The vertical red lines highlight $\phi = 0.5$, which correspond to the slight increases of flux density.
The vertical red lines highlight $\phi = 0.5$ corresponding to the most significant increases of flux density.
%are associated with $\phi \sim$ 0.5 during the decaying giant flare. The dotted vertical lines highlight the zero phases near these flux increases.
}
    \label{fig:flux-orb-phase-2017}
\end{figure*}

\subsection{Search for orbital modulation during the 2017 giant flare}

Since the radio flux emission continuously changes during the 2017 giant flare on hour timescales, it would be speculative to provide a renormalization of the flux density to fold the light curve. Instead, we plotted the 8.5 GHz data, which are the most densely sampled measurements, as a function of the occurrences of zero orbital phases by applying ephemeris from \citet{Bhargava_2017}. Zero orbital phases refer to the X-ray minima at superior conjunction since MJD 40949.384, the $T_{0}$ reference epoch mentioned by \citet{Bhargava_2017}. 
Our first measurement at 8.5 GHz on MJD 57847.438 corresponds to the $84621^{th}$
% $84621.715^{th}$ 
orbital period of Cyg X-3.
For clarity, we reported the occurrences of zero orbital phases since the $84621.0^{th}$ (Cyg X-3 performed about 38 orbital periods in $\sim 8$ days). As shown in Figure~\ref{fig:flux-orb-phase-2017}, all the four enhanced flux densities ranging from $2\,\sigma$ to $6\,\sigma$ during the decaying phase of the flare correspond to $\phi \sim 0.5$, which is aligned with the X-ray maxima. 
The corresponding amplitudes of the peaks range from 2.5\% to 8\% with respect to the general trend of the flare decay.
On the other hand, about $50\%$ of the available measurements at $\phi \sim 0.5$ does not show any significant flux enhancement. Thus, in order to confirm and quantify such a possible phase-flux correlation, a larger and denser monitoring at different frequencies would be required during next giant flares of Cyg X-3.
%For the light-curve described above, the probability of having radio flux enhancements corresponding to $\phi \sim 0.5$$\pm$0.1 by chance is  0.01\%.
%%%\textbf{It would be interesting to investigate on the possible delays and orbital phase studies at the other frequencies by proving a larger and denser monitoring simultaneously at different frequencies during next giant flares of Cyg X-3.}

%and we detect clear variations of
%We performed the timing analysis to search for the 4.8-h orbital period during the 2017 giant flare monitored with Medicina. 
%The folded light curve indicates the maximum of flux close to the 0.5 phase (CHECK), which corresponds to the phase of maximum X-ray... (NORMALIZATION ???; Figure...). 
%(cf paper Miller-Jones giant flare, long-monitoring giant flare?)
%Figure Medicina data with X-ray orbital phases (dashed vertical lines).

\subsection{Orbital modulation during the mini flares}

The light curve obtained during the 2016 mini flare shows the presence of two peaks of different intensity, separated by $\sim 4.8$\,h (see Figure~\ref{fig:6miniflares}). The separation between these peaks
%seems to perfectly
apparently matches with the orbital period of Cyg X-3. We therefore collected all available long and continuous archive observations of several hours
%\textbf{possibly} at least $5$\,h 
obtained during mini-flare episodes to search for a possible orbital modulation during these events. 
Because mini-flares usually last less than a day and are not predictable as giant flares are, only a few observations were carried out by chance. 
%Their studies was made by chance.
% suggest orbital modulation

Different observation programs based on interferometry were conducted in the 1980's in order to study the radio spectral evolution of Cyg X-3 in its low state \citep{Molnar_1984, Molnar_1988}. Mini flares with a flux density $< 1$ Jy were detected during these observations.
%Molnar et al. (1984) studied the spectral evolution of Cyg X-3 in its low state.
The apparent flare separations near the 4.8-h X-ray period suggests a direct correlation with the orbital period.
%Molnar, Reid \& Grindlay (1988) observed a flare of peak flux of $\sim 0.95$ Jy at 22 GHz in 1985 February using VLBI antennas. The observations covered the complete evolution of a flare on $\sim 7$ hours. 
%The source was expanding with a velocity of between 0.16c and 0.312c, again with the expansion preferentially in the north–south direction. 
%
Ten years later, subsequent observations were planned with the VLBA in order to investigate the jet structure during the quiescent state. \citet{Newell_1998} observed two short-duration flares of $\sim 0.3$ Jy and $\sim 0.13$ Jy at 15.3 GHz on 1995 May 7 while the source was in the quiescent state for the rest of the observing run (see Figure~\ref{fig:6miniflares}). Here again, we note a peak separation of $\sim 4.8$ h.
%
%\textbf{We computed the significance of our claimed period detection through simulations. Providing random double peak distributions ($10^8$ trials), the combined probability of chance occurrence of peak separation at the orbital period related to the above datasets is $\sim$0.06\%, even assuming a tolerance of 0.5\,h ($\sim$0.007\% with a 0.25\,h tolerance). Thus, our periodicity claim is statistically significant ($3.5-4.0\,\sigma$ level)}.
%
Thus, among the available data described above, all the datasets
that present a clear double-peak signature also show a peak separation corresponding to the orbital period.

We computed the significance of a possible orbital period detection through simulations
based on the generation of fake light-curves with the same parameters of
the observed data, but introducing random phase shifts for the light-curve peaks.
Providing random double peak distributions (10$^8$ trials), the combined probability
of chance occurrence of peak separation at the orbital period related to the above double-peaked datasets
is $\sim$0.02\%,\,even assuming a conservative tolerance of 20\,min. 
We then included in our statistical analysis  
the remaining datasets 
%additional datasets obtained with the Ryle telescope and the VLA (Figures~\ref{fig:lc-ryle-3june1994} and ~\ref{fig:lc-vla-ryle-25 jan2002}) from \citet{Waltman_1996} and  \citet{Miller-Jones_2009} corresponding to mini-flares of Cyg X-3 
for which no double-peak
distribution seems clearly present or it is poorly significant, assuming no detection for these latter.
The cumulative chance probability (including all data trials) of having only the former datasets
displaying a double-peak distribution at the orbital period is $\sim$0.15\%.
Thus, this preliminary analysis shows that our periodicity claim deserves detailed timing
investigations being statistically significant at $>3$$\sigma$ level.
%
%The combined probability of chance occurrence of peak separation at the orbital period related to the above datasets is $\sim$0.1\%, even assuming a tolerance of 0.5\,h.

%\subsection{Timing tests}

%4 data sets corresponding to the mini-flares, trials with constant $P$, variation on $\dot{P}$, binning, $\chi^{2}$ test, different solutions... 

\begin{figure}
    \centering
    \includegraphics[width=0.4\columnwidth]{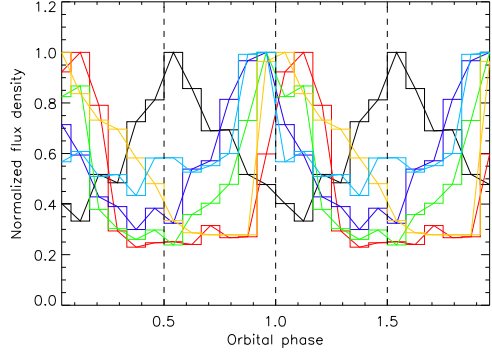}
    \caption{Phase-folded, renormalized light curves corresponding to the 6 mini-flare datasets (green: 1983, blue: 1985, orange: 1994, red: 1995, cyan: 2002, black: 2016), using ephemeris from \citet{Bhargava_2017} and 12 bins. Phase 0 corresponds to the X-ray orbital phase 0 (superior conjunction).}
    \label{fig:folded-4lc}
\end{figure}

In order to investigate periodicities in the multi-year radio light-curves of Cyg X-3, we phase-folded individual datasets at the system's orbital period using the X-ray ephemeris provided by \citet{Bhargava_2017}. Taking into account the number of available measurements, we could provide average flux density values for 12 orbital phase bins of 24 min each. 
%%%We compared the arrival times with and without the timing corrections for the solar system barycenter frame.
%We compared the arrival times with and without the barycentric corrections 
%(\textbf{A: the timing corrections for the solar system barycenter frame}) . 
%%%Because of the orbital period of $4.8$\,h, these barycentric corrections are negligible and they determine phase shifts below $\sim$2\%  level, well within the bin size of our light curves.
In order to ease the comparison among the different datasets, we normalized the folded light curves to their respective peak flux values (see Figure~\ref{fig:folded-4lc}). 
The folded light curves related to $1983-2002$ data are shifted of a half orbital period with respect to the X-ray maxima.
%, but close to the gamma-ray maxima \citep{Fermi_2009} with about 1\,h approximation ($\phi \sim 0.0 \pm 0.1$ in the plot). 
Instead, the 2016 light-curve peaks in correspondence with the X-ray maxima ($\phi \sim 0.5 \pm 0.05$).
%(the peak is around the gamma-ray maxima \citep{Fermi_2009} with about 1\,h approximation ($\phi \sim 0.0 \pm 0.1$ in the plot)
%, while the 2016 light-curve is shifted of a half orbital period, peaking in correspondence with the X-rays maxima \citep[$\phi \sim 0.5 \pm 0.05$;][]{Zdziarski_2012}.

The orbital period is known with very good precision from X-ray observations \citep[$P_{0}=0.19968476(3)$\,d;][]{Bhargava_2017} and cannot be better constrained by the considered radio dataset that includes only a few observed orbital revolutions. 
However, the phase folding of radio data
%the long radio data span (about 33 years) 
could reveal a
different localization of the bulk of the radio emission with respect to the X-ray emission.
The jet emission associated with the 1995 and 1985 mini-flares is seen up to $2.0-2.5$ mas \citep{Newell_1998, Molnar_1988}, while the 2016 mini-flare episode is resolved closer to the compact object %($0.6-0.9$ mas), 
\citep[$0.6-0.9$ mas;][]{Egron_2017}.
Thus the interferometric observations revealed radio emission on the mas scale during mini-flare episodes, 
at distances much larger than the orbital separation ($\sim 3\times10^{11}$\,cm) and the radius of the donor, which is expected to be less than the Roche lobe radius \citep[$\sim\,1.2 \times10^{11}$\,cm;][]{Zdziarski_2018}.
%at a much larger distance from the compact object (the orbital separation being $\sim 3\times10^{11}$\,cm and the radius of the Wolf Rayet star  $\sim 10^{11}$\,cm \citep{Zdziarski_2018}).
%10^{11}$\,cm;][]{Zdziarski_2018}).
This could require 
%additional 
%timing corrections related to the motion and the variable location of the emitting region, 
%in different flaring episodes at different epochs, and therefore it could require 
%additional barycentric 
timing corrections related to the binary system geometry and emitting source motion during different flaring episodes.

Note that the location of the bulk emission and so the amplitude modulation are most likely frequency-dependent, as observed in Cyg X-1 \citep{Brocksopp_2002, Zdziarski_2012b}. Since we  jointly treated the orbital modulation at 15 and 22 GHz, the location of the two emissions during a same flare is expected to be slightly different along the jet.

\citet{Zdziarski_2018} mentioned the presence
%studied the observations of Cyg X-3 obtained at 15 GHz with the Ryle Telescope and the Arcminute Microkelvin Imager (AMI) over 22 yrs. A
of a pronounced peak in the power spectrum of the radio emission at $P_{r} = 0.19944848$\,d, which is shorter by 20\,s than the X-ray orbital period $P_{0}$.
The authors associated this peak with an artefact of the visibility window since it corresponds exactly to fifth harmonic of the sidereal day (frequency at which observations were repeated). An alternative hypothesis (less favored) suggests it may be associated to the presence of a retrograde jet precession with a period of $\sim 170$\,d.
We phase-folded the six individual datasets at $P_{r}$ but the folded light-curves do not appear aligned in phase.
%As shown in Figure~\ref{fig:folded-4lc-20s}, the folded light-curves do not correspond to the same phase.}

% found period with the fifth harmonic of the sidereal day convinced us of its origin as an artefact of the visibility window of the radio telescopes. 

%corresponding to 1983, 1985, and 2016 data are aligned at $\phi \sim 0.5$ while that associated to the 1995 data peaks at $\phi \sim 0.8$. As in the case of the phase-folding at the X-ray orbital period, the four light curves do not correspond to the same phase applying the recent radio orbital period. We then repeated the trials on the period derivative using $P_{r}$ instead of $P_{0}$. In this case, the resulting radio period derivative is $\dot{P} = (1.2 \pm 0.1) \times 10^{-10}$.

%In summary, we have different viable interpretations based on our timing results: considering the standard X-ray ephemeris, (1) the system's radio light curves experienced an actual phase-jump in the time-frame $1995-2016$, or (2) the observed radio emission in 2016 was related to a different emission site/geometry with respect to other mini-flares; (3) the period derivative in radio is different from the X-ray $\dot{P}$ assuming no phase-jumps in the considered data span.
%to X-ray processes, in association with gamma-ray production.

%Figure~\ref{fig:folded-4lc} shows the folded light curves corresponding to the 4 data sets using the ephemeris mentionned previously. 

%\begin{figure*}
%\gridline{\fig{periodogram_P2_mj.eps}{0.4\textwidth}{(a)}
%          \fig{folded-lc-new-pdot-mj.eps}{0.4\textwidth}{(b)}
%          }
%\caption{a) Periodogram of Cyg X-3 where the red dotted line indicates the best radio period derivative while the blue dashed line corresponds to the X-ray period derivative,
%b) phase-folded, renormalized light curve corresponding to the merging of the 6 mini-flare datasets using the best period derivative found from radio data.
%\label{fig:periodogram-foldedlc}}
%\end{figure*}

%\begin{figure}
%    \centering
%    \includegraphics[width=0.4\columnwidth]{folded-6lc-20s.eps}
%    \caption{Phase-folded, renormalized light curves corresponding to the 6 mini-flare datasets (green: 1983, blue: 1985, orange: 1994, red: 1995, cyan: 2002, black: 2016), using ephemeris from \citet{Bhargava_2017} 
%-including the X-ray $\dot{P}$- but $P_{r}$ mentioned by \citet{Zdziarski_2018} and 12 bins. 
%}
%    \label{fig:folded-4lc-20s}
%\end{figure}

\subsection{Discussion}

The X-ray and IR orbital modulations were clearly discovered in 1970's, while
a modulation of the radio emission was suspected at the orbital period in 1980's during the observation of a mini flare of Cyg X-3 \citep{Molnar_1984}. This hypothesis was later discarded looking at long-term radio observations with the Greenbank Interferometer at 2.25 and 8.3 GHz and the Ryle Telescope at 15 GHz. The lack of radio orbital modulation suggested that the bulk of the radio emission originates from a region outside the dense wind of the companion star causing the X-ray and IR scattering \citep{Hjalmarsdotter_2004}, or that the radio emission originating relatively close to the jet base is completely free-free absorbed in the strong wind of the donor \citep{Szostek_2007}.
A recent analysis of the power spectrum associated with the 15-GHz data obtained with the Ryle Telescope and AMI over 22 years has revealed a radio orbital modulation at the orbital period \citep{Zdziarski_2018}.
The modulation varies in amplitude (from $2.5\%$ to $10\%$) and phase based on the X-ray spectral state and radio flux level of Cyg X-3.

We investigated the radio orbital modulation during specific episodes of Cyg X-3: the giant and mini radio flares.
Observations of giant radio flares offer a unique opportunity to investigate the accretion/ejection link under extreme conditions. During such intense emission episodes, large-scale radio jets have been observed with interferometry. VLBA observations performed during the 2001 giant flare showed the expansion of a sequence of individual knots with an initial diameters of 8 mas \citep{Miller-Jones_2004}. The jet precession was found to be over $\sim 5$\,d, significantly lower than during the 1997 flare where the precession period obtained was over 60\,d \citep{Mioduszewski_2001}.
During the decaying emission of the 2017 giant flare, Medicina detected enhanced flux densities that seem to correspond to the orbital phase $\phi \sim 0.5$. 
This suggests that the emission from the base of the jet might be modulated at the orbital period. This could arise most likely from free-free absorption by thermal electrons from the wind of the Wolf Rayet companion.
%Increase of flux could correspond to a new injection of blobs of matter at the base of the jet.
Another scenario would be to assume that there are several jet components that compose the total flare, some of which are launched at the decaying phase of the major flare that could then be susceptible to wind absorption. 
%INCREASE OF FLUX: NEW INJECTION OF BLOBS OF MATTER AT THE BASE OF THE JET $=>$ STRONG ORBITAL MODULATION. THEN MATTER MOVES ALONG THE JET (MORE DISTANT FROM THE COMPACT OBJECT), PHASE SHIFT, NO VISIBLE AT 0.5 ANYMORE ?

%, contrary to the bulk of the unaffected outer jet emission.}

A recent VLBI observation of Cyg X-3 during a mini-flare indicated the presence of two peaks separated by $\sim 4.8$\,h. 
The emission size associated to the flare was found to increase from 0.6 to 0.9 mas (radius) in a few hours \citep{Egron_2017}, corresponding to $\sim$10$^{14}$ cm for a distance of 7.4 kpc. %(see Newell 1998, size).
We note that the light can cross this size in $\sim$1 h, a value compatible with the peak width of the associated light curve shown in Figure~\ref{fig:6miniflares}. Thus, our combination of simultaneous timing and imaging through VLBI observations provides consistent results.

We therefore investigated additional mini-flare observations in order to search for evidence of the radio orbital modulation in Cyg X-3. 
In principle, the modulation of the radio emission should be more pronounced during mini-flares, which correspond to the emission of compact jets closer to the core than transient jets observed during giant flares. Indeed, we expect a higher density of radio-absorbing thermal electrons in this region (from the stellar wind of the companion star and/or from the jet itself) providing a higher opacity and then a stronger orbital modulation. 
%
%We first phase-folded the light curves associated to the mini flares at the X-ray orbital period using the most recent X-ray ephemeris \citep{Bhargava_2017}. Since a shift in phase was clearly visible, we provided trials on the period derivative. A solution is found very close to the X-ray period derivative. We then phase-folded our datasets to the radio orbital period $P_{r}$ mentioned by \citet{Zdziarski_2018} using 22 years of radio data. However, the folded light-curves do not present any peak alignment in phase.
%
We phase-folded the light curves associated with the mini flares at the X-ray orbital period using the most recent X-ray ephemeris \citep{Bhargava_2017}. The 2016 mini-flare is found to be shifted with respect to the other mini-flares ($\phi \sim 0.5$ with respect to $\phi \sim 0$, respectively). The phase shift is most likely attributed to a different location of the radio emitting region along the jet \citep{Egron_2017, Newell_1998, Molnar_1988}. 
%\textbf{Indeed, an emitting jet whose bulk emission is located at an altitude of $10^{14}$ cm above the accretion disk (corresponding to 1 mas angular separation) would provide an emission delayed up to about 1 hour with respect to the compact object
%barycentric 
%position, corresponding to a light-curve phase shift of 0.2.}
%The actual delay depends on the unknown jet location with respect to the accretion disk and the observer line-of-sight.}

The radio orbital modulation could be due to the free-free absorption in the stellar wind of the donor. In this case the minimum flux would be expected at $\phi = 0$, which corresponds to the longest line of sight. However, due to the large amplitude modulation in the light curve, it is difficult to account for such a strong absorption at jet distances much higher than the binary separation. Therefore at least part of the orbital flux modulation could also be attributed to variable/anisotropic shocks in the jet rather than a smooth wind attenuation alone.
Note that \citet{Miller-Jones_2009} arrived at the conclusion that synchrotron self-absorption and free–free absorption by entrained thermal material play a larger role in determining the opacity effects than absorption in the stellar wind of the Wolf Rayet companion. Also \citet{Fender_1997} (based on decreasing opacity of subsequent flares) and \citet{Koljonen_2018} (based on radio spectrum) arrived at the conclusion that the data support thermal plasma mixed with the synchrotron emitting particles that likely means that the wind material is entrained in the jet rather than producing an absorbing screen.
%The phase shift associated with the 2016 mini-flare could be explained by two different hypotheses. In the first one, 

Since the location of the radio emission along the jet can vary,
%corresponding to the 2016 mini-flare is different from the other ones, 
timing correction for the motion and the variable location of the emitting region in different flaring episodes at different epochs would be required to correct for the delay.
%\textbf{When folding the light-curves in our timing analysis, we assumed emission located at the position of the compact object for all the data, since the actual geometry and altitude of the different jet emission episodes is unknown or poorly constrained. In fact, for jets resolved at a few mas, the arrival time delay  of the emission with respect to the compact object position could be significantly high, depending on the actual jet position, extension, inclination and precession. For example, an emitting jet whose bulk emission is located at an altitude of $10^{14}$ cm above the accretion disk (corresponding to 1 mas angular separation) would provide an emission delayed up to about 1 hour with respect to the barycentric position, corresponding to a light-curve phase shift of 0.2. The actual delay depends on the unknown jet geometry with respect to the accretion disk, the absorbing medium and the observer line-of-sight.
%For emission episodes located close to the compact object (altitude less than $10^{13}$cm), no appreciable phase shifts are expected and then light-curves could be coherently folded without further timing correction depending on the jet geometry.}
%
Moreover, jet proper motion as well as jet 
precession
\citep{Dubus_2010, Zdziarski_2018} could also affect the phase shift. For example, a jet ranging $10^{14}$ cm/hour during an observation (as those observed in 1985 and 1995) could provide an additional light-curve phase smearing effect. A better knowledge of the jet geometry, its motion and the absorbing medium would be required to constrain precise time delays and fix related phase corrections. In any case, the observation of orbital modulation is still possible for the relevant height range, since the corresponding phase smearing during a mini-flare event is relatively low even in the unlikely “worst case” of a jet observed during superior/inferior conjunctions (for which timing corrections are at their maximum).
%

%%%However, for jets presenting a certain inclination and extended to distances much larger than the orbital separation (\sim $3\times10^{11}$\,cm),
%when the jet extends to distances much larger than the orbital separation ($3\times10^{11}$\,cm), 
%In the second hypothesis,
%(IPOTESI ZDZIARSKI:) 
Since the jet extends to distances much larger than the orbital separation, it is also possible that 
while the compact object is exactly behind the donor, the radio-emitting region in the bent jet is exactly in front of the donor. In this configuration, the minimum radio flux is at $\phi = 0.5$.
%For jet extensions larger than the orbital separation ($3\times10^{11}$\,cm) and the radius of the Wolf Rayet star  \citep[$\sim 10^{11}$\,cm;][]{Zdziarski_2018}, the radio-emitting region along the jet might be exactly in front of the donor when the compact object is exactly behind the donor. In this configuration, the minimum radio flux is not anymore at $\phi = 0$ but at $\phi = 0.5$. 
Interestingly enough, the 2016 mini-flare emission is found to be closer to the compact object (with respect to the 1985 and 1995 mini-flares), and the flux minimum corresponds to $\phi \sim 0.1$. Instead, the bulk of the radio emission associated with the other mini-flares is located further in the jet and the corresponding flux minimum is shifted of $\sim 0.5$. This is consistent with a bent jet, as shown by \citet{Dubus_2010} and clearly seen by the spatially resolved radio observations. The precession effects are expected to be stronger at a higher distance along the jet.

\citet{Zdziarski_2018} studied the radio orbital modulation based on the X-ray spectral states and radio flux density. The modulation amplitude is found to be higher at low radio fluxes. Clear shifts of the phase of the modulation were observed in the soft, intermediate, and hard states, with a significant dependence on the flux density. The phase of the modulation minimum is found to increase with the increasing radio flux. It can be explained by the distance of the location of the bulk of radio emission increasing with the increasing flux and the jet being inclined with respect to the orbital axis. Interestingly enough, the peak of the radio orbital modulation associated with the data in the highest soft state considered by \citet{Zdziarski_2018} is found at $\phi \sim 0.7$, which is also shifted with respect to the other states ($\phi \sim 0$), and consistent with our results.

%Moreover, jet proper motion as well as jet precession \citet{Dubus_2010, Zdziarski_2018} could also affect the phase shift. For example, a jet ranging $10^{14}$ cm/hour during an observation (as those observed in 1985 and 1995) could provide an additional light-curve phase smearing effect. A better knowledge of the jet geometry, its motion and the absorbing medium would be required to constrain precise time delays and fix related phase corrections. In any case, the observation of orbital modulation is still possible for the relevant height range, since the corresponding phase smearing during a mini-flare event is relatively low even in the unlikely “worst case” of a jet observed during superior/inferior conjunctions (for which timing corrections are at their maximum).}

The peculiarity of the 2016 mini-flare episode could hypothetically reside in the fact that it is seen during the hypersoft X-ray state, 
% \citep{Koljonen_2018,Egron_2017,Trushkin_2017}, a condition possibly related to more compact emission.}
%\textbf{The 2016 mini-flare was observed during the hypersoft X-ray state, 
as clearly observed during the daily monitoring in X-rays with Swift/BAT ($15–50$ keV) and MAXI (2$–$10 keV), and in radio with SMA $(220–230$ GHz), AMI/LA (15 GHz) and RATAN ($4–11$ GHz) \citep{Koljonen_2018, Egron_2017, Trushkin_2017}. The mini-flare was detected quasi-simultaneously to a hard X-ray mini-flare and an enhanced gamma-ray emission, %during the hypersoft X-ray state, 
a few weeks before the giant flare \citep{Egron_2017}. 
Instead, the mini-flares in 1994, 1995 and 2002 did not occur in the hypersoft X-ray states, as shown by the radio/X-ray correlation studied by \citet{Zdziarski_2016} using RXTE/ASM ($1.3-12.2$ keV), MAXI, Swift/BAT, CGRO/BATSE (20–100 keV) and Ryle/AMI (15 GHz). Unfortunately, we do not have information about the X-ray spectral states corresponding to the 1983 and 1985 mini-flares.

During the hypersoft state, the hard X-ray and radio flux densities reach their minima (the jet production is highly diminished or nearly turned off ($1-30$ mJy)) on a similar timescale, while the soft X-ray emission is at its maximum \citep{McCollough_1999}. 
%Mini-flares occurring during the hypersoft states could have a different origin with respect to the other mini-flares. During this state, the hard X-ray flux and radio flux density reach their minima (the jet production is highly diminished or nearly turned off ($1-30$ mJy)) on a similar timescale, while the soft X-ray emission is at its maximum \citep{McCollough_1999}. 
The conditions leading to the formation of the mini-flare during this particular state are likely different from the other ones. In particular, the compact jet appears to be closer to the compact object \citep{Egron_2017}.% during the hypersoft state \citep{Egron_2017}.}

%Indeed, the jet production is highly diminished or nearly turned off ($1-30$ mJy) during the hypersoft X-ray state.

\citet{Fender_1997} suggest that during the radio quenched state (hypersoft X-ray state), the increase of the wind mass-loss rate from the WR star results in an increasing thermal electron density in the region of the compact object, at the location of the base of the jet. This would cause the disappearance of the jet although the exact mechanism remains unclear. 
\citet{Koljonen_2018} propose a scenario to explain the spectral and timing properties of the multi-wavelength observations, and favor the interpretation of changes in the accretion flow instead of changes in the stellar wind structure to explain the accretion state change to/from the hypersoft state. 
They attribute the residual radio emission ($1-30$ mJy) to the wind-wind interaction of the binary and not to the jet.
\citet{Cao_2020} suggest that during the hard to soft X-ray transition, the thin accretion disk extends to the innermost stable circular orbit and replaces the hot accretion flow. This quenches the generation of the internal magnetic field  and causes the magnetic field to quickly diffuse away in the accretion disk, which explains the disappearance of the jets. In this scenario, the subsequent formation of the powerful jets visible during the major and giant radio flares is triggered at a large accretion rate due to the appearance of magnetic outflows above the thin accretion disk, which are able to efficiently advect the strong outer field from the Wolf Rayet.

Also, due to the short orbital distance, the compact object is orbiting the Wolf Rayet star inside the strong wind, which causes interaction between several components - e.g. the jet hitting the wind modulates shocks that dissipate energy effectively. Despite the fast wind speed of Wolf Rayet companion, clumps in the wind \citep{delaCita_2017} could also contribute to a variable and efficient accretion process in addition to a different density in the wind.

\section{Conclusions}

%The study of the radio emission of Cyg X-3 during relativistic ejection periods, in particular during the giant flares, is of utmost important to better understand the link between accretion and ejection.

%The observations of giant flares

Cyg X-3 observations carried out with Medicina at 8.5, 18.6 and 24.1 GHz, in parallel to the MRO at 37 GHz, allowed us to study the evolution of the spectral index along the whole duration of the 2017 giant flare. We confirmed a spectral steepening at the time of the peak maximum of the flare, as observed when previous giant flares had occurred. We highlighted different spectral index variations during the fading of the flare associated with fast changes of the radio emission. A first attempt in the study of the orbital modulation seems to demonstrate that the enhanced radio emissions during the decaying flare might correspond to the orbital phase $\phi \sim 0.5$.
%
%the wind of the Wolf Rayet companion star. 
%The outer jet could instead be affected by major precession effects in the c.
This suggests that the base of the jet could be modulated at the orbital period most likely from free-free absorption by thermal electrons from the wind of the Wolf Rayet companion.
%This suggests that the base of the jet could be modulated at the orbital period most likely by from free-free absorption by thermal electrons either from the wind of the Wolf Rayet companion or from electrons entrained in the jet/plasmon itself.
%

The peak separation of $\sim 4.8$\,h during different single observations most likely indicates an effect of the modulation of the radio emission at the orbital period of the system. Therefore, we systematically searched for radio orbital modulation during mini-flares of Cyg X-3, associated with compact radio emission.
Using the X-ray ephemeris, we found that all the folded light curves appear to be aligned in phase, peaking at $\phi \sim 0$, except for the 2016 light curve which is shifted of $\sim 0.5$ with respect to the other ones. 
%We found a radio timing solution for $\dot{P}$ that provides the alignment in phase of all the light curves. 
%\textbf{We cannot exclude that the apparent period derivative in radio is slightly different from the X-ray $\dot{P}$.}
%because of possible systematic timing errors. 
%In particular, 
The phase shift is likely attributed to a different location of the bulk of the radio emission along the bent jet during the different mini-flares episodes. Time delays arising from jet emission at altitudes $\geq 10^{14}$\,cm can provide significant phase shift $>$\,0.2 (the actual value depending on the jet position and inclination with respect to the observer’s line-of-sight). 
%\textbf{The apparent period derivative in radio is slightly different from the X-ray $\dot{P}$ because of possible systematic timing errors.}
%Indeed, 
The precise location of the radio emitting regions with respect to the accretion disk and the observer is uncertain, even though VLBI observations are shedding some light on the actual dynamic geometry (altitude, extension, line-of-sight inclinations, precession, etc) during the different jet emission episodes.
The shift might also be explained by a particular configuration of the system when the compact object is exactly behind the donor, the radio-emitting region in the inclined jet can be exactly in front of the donor. In this case, the minimum radio flux is at $\phi\,\sim\,0.5$.
In both hypotheses, the phase shift is explained by the large and variable extension of the jets with respect to the orbital separation and radius of the donor.
%The radio orbital modulation could be due to the free-free absorption in the stellar wind of the donor and a variable/anisotropic shocks in the jet.
The observed large depths in the folded light curves put in doubt the possibility that the orbital modulation could be fully attributed to free-free absorption in the stellar wind if the emission is occurring at large distances from the binary system. A contribution of variable/anisotropic shocks in the jet could explain at least part of the modulation.

Our timing results  might suggest that the studied mini-flares occurred during different accretion states and/or different emission geometries.
%In fact, the peculiar 2016 mini-flare appeared during a radio-quenched hypersoft X-ray state, and it showed a 0.5 phase shift compared to the other mini-flare episodes. 
Furthermore, the different peak widths and light curve shapes observed during mini-flare episodes could be explained as phase smearing effects related to jet proper motion and precession. In extreme cases this could also prevent orbital period detection, if for example the proper motion along the line-of-sight would exceed $5\times10^{14}$\,cm/hour.
%1) the studied mini-flares occurred within different emission geometries to be better constrained by further VLBI observations;
%2) 

%\textbf{Based on our results, we proposed different viable (independent or combined) interpretations that can be summarized as follows: 
%1) the studied mini-flares occurred within different emission geometries to be better constrained by further VLBI observations;
%2) the studied mini-flares occurred during different accretion states; in fact the peculiar 2016 mini-flare occurred during a radio-quenched hypersoft X-ray state, and it showed a 0.5 phase shift compared to the other mini-flare episodes;}
%2) the period derivative in radio is different from the X-ray $\dot{P}$; 3) the jet precession varies according to the jet velocity and extension; 4) the system's radio light curves experienced slight phase drifts. 
%
%{2) the apparent period derivative in radio is slightly different from the X-ray $\dot{P}$ (possibly due to different emission regions); 
%3) the system’s radio light curves experienced slight phase drifts, hypothetically related to the timing irregularities recently detected in X-rays \citep{Bhargava_2017, Antokhin_2019}.
% and/or to the jet precession \citep{Zdziarski_2018}}.
%
Additional observations of mini-flares in radio, X-rays and gamma-rays during different states of the source will help to shed light on the phase shift observed in our present data sets. %and confirm or rule out these hypotheses.
In particular, further VLBI observations are needed to better constrain the emission geometry also required for coherent timing analysis.

%The different hypotheses or combination of hypotheses to explain the shifts in phase seen during the mini flares can be summarized as follows: 1) the period derivative in radio is different from the X-ray $\dot{P}$; %2) mini-flares occur during different states of the source that are associated with different accretion states and emission site/geometry; 3) the jet precession varies according to the jet velocity and extension; %4) the system's radio light curves experienced slight phase drifts.

%Continuous changes in radio, precession, jets...

\section*{Acknowledgements}
%Authors from VLBI (ask Marcello) ?
%M.P. acknowledges financial support from the RAS (CRP-25476).
%E.K. acknowledges support from TUBITAK Grant 115F488.
We thank the referee for her/his constructive suggestions, which improved the discussion of the paper.
Based on observations with the Medicina telescope operated by INAF - Istituto di Radioastronomia. We thank A. Zanichelli for promptly scheduling the Medicina observations.
The Sardinia Radio Telescope is funded by the Department of University and Research (MIUR), the Italian Space Agency (ASI), and the Autonomous Region of Sardinia (RAS), and is operated as a National Facility by the National Institute for Astrophysics (INAF). The Torun radio telescope is operated by Torun Centre for Astronomy of Nicolaus Copernicus University in Torun (Poland) and supported by the PolishMinistry of Science andHigher Education SpUBgrant.
S.C. acknowledges the financial support from the UnivEarthS Labex program of Sorbonne Paris Cit\'{e} (ANR-10-LABX-0023 and ANR-11- IDEX-0005-02). 
V.G. is supported through the Margarethe von Wrangell fellowship by the European Social Fund and the Ministry of Science, Research and the Arts Baden-W\"{u}rttemberg. JR ackowledges partial funding from the French space agency (CNES).

\appendix

\section{Flux density measurements during the giant flare of April 2017}

In the following, we present the data associated to the giant flare of Cyg X-3 in April 2017, obtained with Medicina at 8.5, 18.6 and 24.1 GHz (Tables~\ref{tab:A1}, ~\ref{tab:A2} and ~\ref{tab:A3}, respectively), and with the Mets\"ahovi radio observatory (MRO) at 37 GHz (Table~\ref{tab:A4}). The Tables~\ref{tab:A1}, \ref{tab:A2} and \ref{tab:A3} are published in their entirety in the machine-readable format. A portion is shown here for guidance regarding its form and content. The data are shown in the light curves presented in Figure~\ref{fig:flux-alpha-2017}.

\begin{deluxetable*}{ccc}
\tablenum{A1}
\tablecaption{Medicina data obtained at 8.5 GHz from 4 to 11 April 2017\label{tab:A1}.}
\tablewidth{0pt}
\tablehead{
\colhead{MJD} & \colhead{Flux} & \colhead{Errors} \\
\colhead{} & \colhead{(Jy)} & \colhead{(Jy)}
%& \multicolumn2c{(kpc)} & \colhead{Constellation} & \colhead{(mag)}
}
%\decimalcolnumbers
\startdata
%M1 & NGC 1952 & Crab Nebula & Supernova remnant & 2 & Taurus & 8.4 \\
 57847.43834 &  5.2823 &   0.0745  \\
 57847.44047 &  5.3301 &   0.0757  \\
 57847.44259 &  5.4555 &   0.0708  \\
 57847.44471 &  5.5285 &   0.0844  \\
 57847.44684 &  5.5964 &   0.0818  \\
 57847.44897 &  5.6221 &   0.0774  \\
 57847.50248 &  6.8600 &   0.0779  \\
 57847.50459 &  6.9215 &   0.0901  \\
 57847.50671 &  7.0066 &   0.0893  \\
 57847.50883 &  7.1401 &   0.0889  \\
\enddata
%\tablecomments{Table 1 is published in its entirety in the machine-readable format. A portion is shown here for guidance regarding its form and content.}
\end{deluxetable*}

\begin{deluxetable*}{ccc}
\tablenum{A2}
\tablecaption{Medicina data obtained at 18.6 GHz from 5 to 11 April 2017\label{tab:A2}.}
\tablewidth{0pt}
\tablehead{
\colhead{MJD} & \colhead{Flux} & \colhead{Errors} \\
\colhead{} & \colhead{(Jy)} & \colhead{(Jy)}
%& \multicolumn2c{(kpc)} & \colhead{Constellation} & \colhead{(mag)}
}
%\decimalcolnumbers
\startdata
%M1 & NGC 1952 & Crab Nebula & Supernova remnant & 2 & Taurus & 8.4 \\
57848.062  &  13.875 & 0.348 \\
57848.064  &  13.637 & 0.337 \\
57848.066  &  13.803 & 0.332 \\
57848.068  &  13.961 & 0.336 \\
57848.069  &  13.837 & 0.332 \\
57848.122  &  13.588 & 0.238 \\
57848.124  &  13.581 & 0.232 \\
57848.126  &  13.503 & 0.228 \\
57848.128  &  13.000 & 0.388 \\
57848.130  &  12.467 & 0.236 \\
\enddata
%\tablecomments{Table 1 is published in its entirety in the machine-readable format. A portion is shown here for guidance regarding its form and content.}
\end{deluxetable*}

\begin{deluxetable*}{ccc}
\tablenum{A3}
\tablecaption{Medicina data obtained at 24.1 GHz from 4 to 11 April 2017\label{tab:A3}.}
\tablewidth{0pt}
\tablehead{
\colhead{MJD} & \colhead{Flux} & \colhead{Errors} \\
\colhead{} & \colhead{(Jy)} & \colhead{(Jy)}
%& \multicolumn2c{(kpc)} & \colhead{Constellation} & \colhead{(mag)}
}
%\decimalcolnumbers
\startdata
%M1 & NGC 1952 & Crab Nebula & Supernova remnant & 2 & Taurus & 8.4 \\
57847.464 &   3.395 & 0.112 \\
57847.466 &   3.398 & 0.117 \\
57847.469 &   3.402 & 0.126 \\
57847.471 &   3.384 & 0.124 \\
57847.473 &   3.324 & 0.118 \\
57847.475 &   3.366 & 0.122 \\
57847.477 &   3.692 & 0.156 \\
57847.479 &   3.707 & 0.152 \\
57847.482 &   3.756 & 0.149 \\
57847.484 &   4.013 & 0.161 \\
\enddata
%\tablecomments{Table 1 is published in its entirety in the machine-readable format. A portion is shown here for guidance regarding its form and content.}
\end{deluxetable*}

\begin{deluxetable*}{ccc}
\tablenum{A4}
\tablecaption{Mets\"ahovi radio observatory data obtained at 37 GHz from 4 to 16 April 2017\label{tab:A4}. For clarity, only the data from 4 to 11 April are shown on Figure ~\ref{fig:flux-alpha-2017}.}
\tablewidth{0pt}
\tablehead{
\colhead{MJD} & \colhead{Flux} & \colhead{Errors} \\
\colhead{} & \colhead{(Jy)} & \colhead{(Jy)}
%& \multicolumn2c{(kpc)} & \colhead{Constellation} & \colhead{(mag)}
}
%\decimalcolnumbers
\startdata
%M1 & NGC 1952 & Crab Nebula & Supernova remnant & 2 & Taurus & 8.4 \\
57847.410 &	2.2644 &	0.2059 \\
57848.076 &	10.3995 & 	0.3518 \\
57848.404 &	6.9656 &	0.3531 \\
57849.115 &	6.4108 &	0.2047 \\
57849.140 &	7.3071 &	0.2346 \\
57849.347 &	6.8063 &	0.2294 \\
57850.051 &	5.3973 &	0.1782 \\
57850.104 &	6.2294 &	0.2588 \\
57850.329 &	5.6663 &	0.2523 \\
57850.349 &	5.7361 &	0.2434 \\
57850.390 &	5.6497 &	0.3567 \\
57850.429 & 6.3364 &	0.3796 \\
57850.455 & 6.6414 &	0.2842 \\
57851.069 &	4.4258 &	0.1637 \\
57851.092 &	3.7249 &	0.1273 \\
57851.132 &	4.7308 &	0.1796 \\
57851.342 &	2.9154 &	0.2042 \\
57851.364 &	3.4432 &	0.154 \\
57851.406 &	3.3389 &	0.2358 \\
57851.428 &	3.8986 &	0.1662 \\
57852.080 &	2.2999 &	0.1088 \\
57852.102 &	2.3946 &	0.0953 \\
57852.124 &	2.5624 &	0.1035 \\
57852.349 &	1.755  &	0.1068 \\
57852.402 &	1.8095 &	0.136 \\
57853.050 &	1.5108 &	0.096 \\
57853.072 &	1.3854 &	0.0853 \\
57853.114 &	1.4414 &	0.1015 \\
57854.351 &	0.6786 &	0.1861 \\
57855.059 & 0.6109 &    0.0719  \\
57857.046 & 0.4702 &    0.0839  \\
57858.061 & 0.154  &    0.0782  \\
57859.044 & 0.5408 &    0.0736 \\
\enddata
%\tablecomments{Table 1 is published in its entirety in the machine-readable format. A portion is shown here for guidance regarding its form and content.}
\end{deluxetable*}

%%%%%%%%%%%%%%%%%%%%%%%%%%%%%%%%%%%%%%%%%%%%%%%%%%

%%%%%%%%%%%%%%%%%%%% REFERENCES %%%%%%%%%%%%%%%%%%

%% For this sample we use BibTeX plus aasjournals.bst to generate the
%% the bibliography. The sample63.bib file was populated from ADS. To
%% get the citations to show in the compiled file do the following:
%%
%% pdflatex sample63.tex
%% bibtext sample63
%% pdflatex sample63.tex
%% pdflatex sample63.tex

\bibliography{CygX3}{}
\bibliographystyle{aasjournal}

%% This command is needed to show the entire author+affiliation list when
%% the collaboration and author truncation commands are used.  It has to
%% go at the end of the manuscript.
%\allauthors

%% Include this line if you are using the \added, \replaced, \deleted
%% commands to see a summary list of all changes at the end of the article.
%\listofchanges

\end{document}